\documentclass[showpacs,preprintnumbers,amsmath,amssymb,floatfix,12pt]{revtex4}
\usepackage{graphicx}
\usepackage{dcolumn}
\usepackage{bm}

\begin{document}

%

\let\a=\alpha      \let\b=\beta       \let\c=\chi        \let\d=\delta
\let\e=\varepsilon \let\f=\varphi     \let\g=\gamma      \let\h=\eta
\let\k=\kappa      \let\l=\lambda     \let\m=\mu
\let\o=\omega      \let\r=\varrho     \let\s=\sigma
\let\t=\tau        \let\th=\vartheta  \let\y=\upsilon    \let\x=\xi
\let\z=\zeta       \let\io=\iota      \let\vp=\varpi     \let\ro=\rho
\let\ph=\phi       \let\ep=\epsilon   \let\te=\theta
\let\n=\nu
\let\D=\Delta   \let\F=\Phi    \let\G=\Gamma  \let\L=\Lambda
\let\O=\Omega   \let\P=\Pi     \let\Ps=\Psi   \let\Si=\Sigma
\let\Th=\Theta  \let\X=\Xi     \let\Y=\Upsilon

%

%

\def\cA{{\cal A}}                \def\cB{{\cal B}}
\def\cC{{\cal C}}                \def\cD{{\cal D}}
\def\cE{{\cal E}}                \def\cF{{\cal F}}
\def\cG{{\cal G}}                \def\cH{{\cal H}}
\def\cI{{\cal I}}                \def\cJ{{\cal J}}
\def\cK{{\cal K}}                \def\cL{{\cal L}}
\def\cM{{\cal M}}                \def\cN{{\cal N}}
\def\cO{{\cal O}}                \def\cP{{\cal P}}
\def\cQ{{\cal Q}}                \def\cR{{\cal R}}
\def\cS{{\cal S}}                \def\cT{{\cal T}}
\def\cU{{\cal U}}                \def\cV{{\cal V}}
\def\cW{{\cal W}}                \def\cX{{\cal X}}
\def\cY{{\cal Y}}                \def\cZ{{\cal Z}}

%

\newcommand{\Ns}{N\hspace{-4.7mm}\not\hspace{2.7mm}}
\newcommand{\qs}{q\hspace{-3.7mm}\not\hspace{3.4mm}}
\newcommand{\ps}{p\hspace{-3.3mm}\not\hspace{1.2mm}}
\newcommand{\ks}{k\hspace{-3.3mm}\not\hspace{1.2mm}}
\newcommand{\des}{\partial\hspace{-4.mm}\not\hspace{2.5mm}}
\newcommand{\desco}{D\hspace{-4mm}\not\hspace{2mm}}

\def\deltaakpi{\Delta {{A}}_{K\pi}}


\title{\boldmath 
Interpreting $Z(3900)$ }

\author{Namit Mahajan
}
\email{nmahajan@prl.res.in}
\affiliation{
 Theoretical Physics Division, Physical Research Laboratory, Navrangpura, Ahmedabad
380 009, India
}


\begin{abstract}
A new charged state, called $Z_c(3900)$ has been recently observed by BESIII collaboration in $J/\psi\pi^{\pm}$ channel. The same has been confirmed by Belle (called $Z(3894.5)$ in the Belle paper). In this short note we discuss probable interpretations of this charmonium like state which we call $Z(3900)$. 

\end{abstract}

\pacs{
 }
\maketitle


A large number of states have been observed during the last decade or so which can not be accommodated within the quark model \cite{exotica}. BESIII has reported a charged state 
in the $J/\psi\pi^{\pm}$ channel in $Y(4260)\to J/\psi\pi^+\pi^-$ decay \cite{Ablikim:2013mio}. The reported mass and width of the new charged state, called $Z_c(3900)$ are $3899.0 \pm 3.6 \pm 4.9$ MeV and $46 \pm 10 \pm 20$ MeV respectively with a significance more than $8\sigma$. Belle  \cite{Liu:2013dau} has also reported a charged state, called $Z(3895)$, with mass $3894.5 \pm 6.6 \pm 4.5$ MeV and width $63 \pm 24 \pm 26$ MeV observed with a significance of more thn $5\sigma$ in the $J/\psi\pi^{\pm}$ distribution in exactly the same decay channel of $Y(4260)$ as in the BESIII analysis. The Belle observation supports the claim made by BESIII for the existence of a charmonium like charged state. Making a naive combination of these two results yields $M = 3897.4 \pm 4.8$ MeV and $\Gamma = 50.9 \pm 18.9$ MeV. We call this state $Z(3900)$.
It joins the family of other quarkoniun like charged states: $Z^+(4430)$ which was observed as a peak in the $\psi'\pi^{\pm}$ channel \cite{Choi:2007wga} and $Z_1(4051)$, $Z_2(4248)$ both of which decay to $\chi_{c_1}\pi^{\pm}$ \cite{Mizuk:2008me}. Earlier, similar charged bottomonium states were reported by Belle \cite{Belle:2011aa}: $Z_b(10610)$ decaying to $\Upsilon(nS)\pi^{\pm}\,\,(n=1,2,3)$ and $Z'_b(10650)$ decaying into $h_b(mP)\pi^{\pm}\,\,(m=1,2)$. All these charged states have been observed/discovered in the processes $e^+e^-\longrightarrow [\bar{Q}Q] \pi^+\pi^-$, where $Q = c, b$. Due to proximity of these states with the corresponding $H^{(*)}\bar{H^{(*)}}$ states (where $H^{(*)}$ refers to a generic charm or bottom meson), it is always tempting to identify them as weakly bound molecular states of heavy mesons. However, as we discuss below, molecular interpretation may run into trouble.

Being a charged state which decays to a charmonium and a pion, $Z^+(4430)$ can certainly be not a member of the conventional charmonium family in any way and also can not be a hybrid charmonium state. Assuming an S-wave Breit-Wigner fit for the $J/\psi \pi$ spectrum , $Z(3900)$ would have $I^G(J^P)$ quantum numbers as $1^+(1^+)$. 
A neutral state, the isospin neutral partner, should exist and would have a mass differing by a few MeVs with $C = -1$.
Both BESIII and Belle  parametrize the signal shape as an S-wave Breit-Wigner function. BESIII also performs a fit by assuming a P-wave between $Z_c(3900)$ and $\pi$ and also between $J/\psi$ and $\pi$. In the latter case, the parity assignment would have to be opposite. 
The mass of $Z(3900)$ is way above the $D^+\bar{D^0}$ threshold but is only about $23$ MeV more than $D^{*+}\bar{D^0}$ threshold. Such a possibility has been studied in \cite{Wang:2013cya}.
Though this assignment would easily fit the assumed quantum numbers above with an S-wave molecular bonding, it is difficult to reconcile with a large positive mass difference between the bound state and the constituents. It is rather like an excitation energy. However, a non-zero relative angular momentum between $D$ and $D^*$ could possibly sustain a bound state due to the angular momentum barrier. If the parity does finally turn out to be positive, then odd $\ell$ values for the relative angular momentum would be ruled out. Furthermore, such a state will have a tendency to decay into $D\bar{D}\pi$ and $D\bar{D}\gamma$.
The molecular state formed out of $D^{*+}\bar{D^*}$ will have a binding energy in excess of $100$ MeV which implies a size ${\mathcal{O}}(0.1)$ fm. In such a case, it may not be correct to consider only the $D^{*+}\bar{D^*}$ component, which may have a relatively small weight compared to the core component. 
Moreover, for molecular states with hidden flavour, there is a possibility of annihilating into ordinary mesons. Using the one pion exchange potential (OPEP) \cite{Shmatikov:1995td}
estimates the decay width in terms of the binding energy $E_B$:
\[
\Gamma \sim E_B \frac{4}{3}\sqrt{\frac{m_Q}{E_B}} \left(\alpha_s \frac{1}{r_{eff} m_Q}\right)^2
\]
where $r_{eff}$ denotes the range/radius of the binding force. Choosing $E_B$ to be that for the $D^{*+}\bar{D^0}$ molecular state and $r_{eff} = \frac{1}{4m_{\pi}}$,
one finds that $\Gamma \sim {\mathcal{O}}(10)$ MeV while if one chooses $r_{eff} = \frac{1}{2m_{\pi}}$, the width turns out to be only a few MeV. This argument will perhaps not hold when a sizeable core component is included as OPEP will not suffice and there will be a short range force responsible for such a tight binding. However, in that case it will not be correct to call it a molecular state. 
We therefore conclude that in the most probable case, $Z(3900)$ can not be reliably described as a molecular state. 

Next consider the tetraquark as a possibility for the new found state. There are many possibilities in which a diquark combines with antidiquark. This leads to a large number of states possible. Here we consider the following two: $[cc\bar{u}\bar{d}]$ and $[\bar{c}cu\bar{d}]$. The former leads to $D^{(*)+}D^0$  while the later to $D^{(*)+}\bar{D^0}$ structure after rearrangement. 
Tornqvist \cite{Tornqvist:1991ks} has argued that the binding in $I = 0$ channel for $D^*D$ is thrice as strong as that in $I = 1$ channel. It is therefore important to search for such isoscalar states which would decay to say $J/\psi\eta$. Moreover, neglecting the difference in the masses of up and down quark masses (isospin symmetry), there should be degenerate states in each channel. Also considering OPEP as a source of binding, it is generally argued that only one of the structures can bind. The argument relies on the fact that since the vertices involve the third component of isospin, if the potential is attractive for one, it is repulsive for the other. This however may not be a correct expectation in reality, and care must be exercised. 
\cite{Ebert:2007rn} shows that the $[cc\bar{q}\bar{q'}]$ tetraquark states have masses above the thresholds for decays to open charm while \cite{Ebert:2005nc} has shown that S-wave hidden flavour tetraquarks in the charm sector typically lie above the corresponding open charm thresholds, a property distinct from the bottomonium sector. The reason is that the attractive colour force is the same but the repulsive kinetic energy is much smaller in the bottomonium compared to charmonium. In particular, the state with $J^{PC} = 1^{+-}$ has a predicted mass of $3890$ MeV (compare this with the same state with predicted mass of $3882$ MeV in \cite{Maiani:2004vq}). Ref.\cite{Faccini:2013lda} discusses the  tetraquark and molecular possibilities for the $Z(3900)$ state and have provided estimates for various branching fractions. 
In any case, either of the tetraquark states above is heavier than sum of the masses of the constituents, and there should be a tendency to quickly fall apart into the constituent mesons, though for the hidden charm state decays to $J/\psi\pi$ are quite probable. The branching fraction into $J/\psi\pi$ may turn out to be smaller than estimated. 
The conclusion to be drawn from the above discussion is that the tetraquark remains a viable possibility for the newly found state $Z(3900)$.

We now turn to the possibility of identifying $Z(3900)$ with one of the hadro-charmonium states. Hadro-charmonium (more generally hadro-quarkonium) \cite{Dubynskiy:2008mq}
 was suggested as a possible option for the charged state $Z^+(4430)$ which despite being above various open charm thresholds had an affinity to go into $\psi'\pi^+$ (the decay into $J/\psi\pi^+$ is suppressed). The main idea behind a hadro-quarkonium is that the particular quarkonium is assumed to be embedded in light hadronic environment. This puts a hadro-quarkonium from molecular and other multi-quark states. The interaction between the quarkonium and the light hadronic medium around it is the QCD van der  Waals force. See for example \cite{Brodsky:1989jd}-\cite{Fujii:1999xn} for various aspects of quarkonium interactions with ordinary matter.
The affinity towards a particular quarkonium state is partly due to the fact that the corresponding quarkonium state has the two heavy quarks in a specific spin state. This specific spin state is just sitting in the hadronic environment in such a way that the interaction with the environment does not destroy the core properties of the quarkonium. This reasoning is motivated from the conclusion reached (see Kharzeev and Satz in \cite{Brodsky:1989jd}) about the quarkonia not being broken up in hadronic matter, though mass shifts are perfectly allowed. Various estimates suggest that for $J/\psi$ the binding energy in ordinary nuclear matter is $\sim 10$ Mev while for $\Upsilon$ it is few MeVs. For $\psi'$ the naive estimate turns out to be anywhere between $100$ MeV (see Lee and Ko in \cite{Brodsky:1989jd}) and $700$ MeV \cite{Luke:1992tm}. These are taken to be only indicative values only and not to be taken too literally since one of the implicit assumptions in all these estimates is the applicability of multipole expansion and its convergence. More support to the hadro-quarkonium emerges from the computation of long range force between two small colour dipoles (small colour dipoles realise themselves as $\bar{Q}Q$ states \cite{Fujii:1999xn}. It is found that the long range interactions between two such dipoles are dominated by pion clouds extending to $(2m_{\pi})^{-1}$. The quarkonium size is $\sim (\alpha_s m_Q)^{-1}$ and the range of the gluon (light hadronic) matter will be $E_B^{-1} \sim (\alpha_s^2 m_Q)^{-1}$. In this picture, the quarkonium, a compact colour singlet objet, is embedded in a gluonic environment which is then surrounded by a pionic cloud. This therefore lends direct support to the above ideas. 

Building on the arguments presented above, we suggest that all the charged charmonium like states observed i.e. $Z(3900)$, \{$Z_1(4051)$, $Z_2(4248)$\} and $Z(4430)$ are hadro-charmonium states containing $J/\psi$, $\chi_{c_1}$ and $\psi'$. The difference in masses of the individual charmonia are close to the difference in masses of the suggested hadro-charmonia. One can therefore think of these as possible excited states, whenever applicable as in the case of $J/\psi$ and $\psi'$. This picture would be confirmed if for example there is another state, which we call $Z(4500)$, at around $4500$ MeV or so decaying dominantly to $\psi(3770)\pi$. Also and more importantly, there should be a state that contains $\eta_c$. Taking a clue from the pattern above, the expected mass of such a state would be about $3800$ MeV. Till now, the discussion of hadro-charmonium has assumed ordinary matter but in reality the strange matter may also be present. In that case, corresponding states decaying into charmonia and a kaon should exist. However, strange quark is not as light as are up and down quarks and therefore there may be significant differences. This should be explored in detail though within the chiral theory both kaons and pions are treated as Goldstone bosons. But it is known that $SU(3)$ breaking effects could be appreciable. It is likely that the $Z_b(10610)$ and $Z'_b(10650)$ also fall in this category of states though each shows transition to various $b\bar{b}$ states. However, as noted in \cite{Belle:2011aa}, the angular analysis indicates $J^P = 1^+$ and G-parity $+$ for these states. This makes it very plausible that all the charged quarkonium like states observed are hadro-quarkonia.

In this short note we have looked at various possible options for the newly discovered state $Z(3900)$ by BESIII and Belle collaborations. We have argued that the molecular interpretation is less likely. Identification as a tetraquark state is a viable option but it is not very clear if decay to open charm states is actually suppressed since the state lies appreciably above the open charm threshold. This could be investigated with QCD sum rules and we leave it for future study. Hadro-charmonium is a tempting option as it readily explains affinity to a particular charmonium state. We have suggested that it is highly likely that all the charged exotic states, observed both in the charm and bottom sectors, are in fact hadro-quarkonia. Observation of states containing $\eta_{c,b}$ and $\psi(3770)$ will be a clear confirmation of this picture. It is however very important to measure the branching fraction into open charm states like $\bar{D}D^{*+}$ to reach any definite conclusion. This could be motivated by noting that $\psi(3770)$ has a decay width of about $25$ MeV and lies approximately $35$ MeV above the $D\bar{D}$ threshold. The decay into $D\bar{D}$ indeed turns out to be dominant one for this state. There is a high degree of similarity between this example and $Z(3900)$ and $D^{*+}\bar{D}$ threshold. More information on other decay modes and experimental determination of quantum numbers will be crucial in pin pointing the real interpretation. Before closing, we would like to mention that $Z(4430)$ and $Z_1,\,\, Z_2$ were observed in B-decays to two pions and respective charmonia. Similar search should be made in $B \longrightarrow J/\psi \pi^+\pi^-$ mode (PDG \cite{Beringer:1900zz} quotes a branching ratio $(4.6 \pm 0.9) \times 10^{-5}$ for this mode which is similar to that for $\psi'$ and $\chi_{c_1}$ modes). What may be more fruitful would be to look at $B \longrightarrow J/\psi K^+\pi^-$ as this has a branching fraction which is two orders of magnitude higher than $B \longrightarrow J/\psi \pi^+\pi^-$. Dalitz plot analysis in $M_{K\pi}$ vs $M_{J/\psi\pi}$ should easily reveal such a state. In fact, Dalitz plot analysis could be done for all the three possible pairs which will unambigouosly reveal the presence of any underlying resonant structure.
This observation i.e. looking at B-decays with a specific charmonium and a kaon plus one (or two) pion(s) in the final may be fruitful for searching and confirming other states, including states with strangeness, since typically these have branching fractions which are at least one order of magnitude higher than those with two pions states. It may also be of interest and relevance to look for states with a charmonium and two kaons in the final state. In this connection, a decay like $\Upsilon(5S) \longrightarrow \Upsilon(1S) K^+ K^-$ may play a crucial role in testing various ideas.

\vskip 0.5cm
{\it Note}: As this manuscript was being finalised, there appeared \cite{Voloshin:2013dpa} and \cite{Karliner:2013dqa}. \cite{Voloshin:2013dpa} discusses molecular and hadro-charmonium as possible options for $Z_c(3900)$ while \cite{Karliner:2013dqa} analyses doubly heavy tetraquark and baryonic states in general with comments and remarks relevent for the new state.



%

\end{document}